\documentclass[prl,twocolumn]{revtex4}
\usepackage{graphicx, epsfig}
\usepackage{color}
\usepackage{mathrsfs}
\usepackage{bm}
\usepackage{amsmath,amssymb,yhmath}
\usepackage[caption=false]{subfig}
\usepackage{hyperref}
\usepackage[normalem]{ulem}
\usepackage{mathtools}

\newcommand{\be}{\begin{equation}}
\newcommand{\ee}{\end{equation}}
\newcommand{\ba}{\begin{eqnarray}}
\newcommand{\ea}{\end{eqnarray}}

\begin{document}
\title{Decay of Cosmic String Loops Due to Particle Radiation}
\author{Daiju Matsunami$^\circ$$^\dag$, Levon Pogosian$^\dag$, Ayush Saurabh$^*$, Tanmay Vachaspati$^*$}
\affiliation{
$^\circ$Canadian Institute for Theoretical Astrophysics, University of Toronto, Toronto, ON M5S 3H8, Canada \\
$^\dag$Physics Department, Simon Fraser University, Burnaby, BC V5A 1S6, Canada. \\
$^*$Physics Department, Arizona State University, Tempe, AZ 85287, USA. \\
}

\begin{abstract}
\noindent
Constraints on the tension and the abundance of cosmic strings depend crucially on the rate at which they decay into particles and gravitational radiation. We study the decay of cosmic string loops in the Abelian-Higgs model by performing field theory simulations of loop formation and evolution. We find that our set of string loops emit particle radiation primarily due to kink collisions, and that their decay time due to these losses is proportional to $L^p$ with $p \approx 2$ where $L$ is the loop length. In contrast, the decay time to gravitational radiation scales in proportion to $L$, and we conclude that particle emission is the primary energy loss mechanism for loops smaller than a critical length scale, while gravitational losses dominate for larger loops.
\end{abstract}

\maketitle

Cosmic strings play an important role in building theories of the early universe \cite{Vilenkin:2000jqa} and provide a rare observational probe of String Theory \cite{Copeland:2011dx}. The search for their signatures has mostly focused on their gravitational effects, and they are among the main science goals of LIGO \cite{Abbott:2017mem}. The tightest bound on the string tension $\mu$, coming from millisecond pulsar timing measurements \cite{Lasky:2015lej}, is based on the gravitational wave (GW) background produced by decaying cosmic string loops. This bound, $G\mu \lesssim 10^{-10}$ \cite{Blanco-Pillado:2017rnf,Abbott:2017mem}, where $G$ is Newton's gravitational constant, assumes that string loops decay primarily into GW with the quantitative predictions obtained from simulations using the Nambu-Goto (NG) approximation that ignores the field composition of the strings \cite{Albrecht:1984xv,Bennett:1989yp,Allen:1990tv,BlancoPillado:2011dq,Lorenz:2010sm}. While it is widely accepted that the NG description works well for loops much larger than the string width, the exact loop size above which the particle composition of the string cores can be ignored is not firmly established. The few existing field theory simulations of string networks suggest that loops primarily decay into particle radiation~\cite{Hindmarsh:2017qff}, with cosmological size loops not surviving beyond one oscillation, potentially leading to a new paradigm for cosmic string evolution in which the GW bounds do not apply. Thus it is critical to examine particle emission by cosmic string loops and to determine their primary decay mode.

Previous studies of the particle radiation from cosmic strings included analytical 
estimates~\cite{Vachaspati:1984yi}, some based on effective couplings of NG 
strings to other fields~\cite{Srednicki:1986xg,Brandenberger:1986vj}, 
field theory simulations of standing waves, 
kinks and cusps on long strings \cite{Olum:1998ag,Olum:1999sg} and simulations 
of strings with small oscillations~\cite{Martins:2003vd,Hindmarsh:2017qff}.
In this Letter, for the first time, we directly examine the decay of a cosmic string loop 
to particle radiation in the Abelian-Higgs model by simulating 
loop formation followed by evolution in full field theory. 
The focus on a single loop is to be contrasted with the very large field theory simulations
of an entire network of strings in an expanding spacetime~\cite{Hindmarsh:2008dw,Hindmarsh:2017qff}. 

We find that string loops emit particle radiation mainly due to features on the strings known as kinks and cusps
\cite{Vilenkin:2000jqa}. The half-life of a loop due to particle radiation is proportional to 
$L^p$, where $L$
is the length of the loop and $p\approx 2$ for the loops we have considered.
On the other hand, the loop half-life due to gravitational
radiation is known to be proportional to $L$. Thus, there is a crossover from
particle-decay to gravitational-decay roughly
given by $L_* \sim w/G\mu$ where $w\sim \mu^{-1/2}$ is the width of the string. For $L < L_*$,
loops decay by particle emission, while for $L > L_*$ gravitational emission dominates.
We discuss caveats and the implications of this result in more detail below, along with 
the values of $p$ that might arise for loops other than
those we have directly simulated.

We consider the Abelian-Higgs field theory with a complex scalar field, $\phi =\phi_1+i\phi_2$,
and a U(1) gauge field, $A_\mu$. We work in the temporal gauge, $A_0=0$, and
the field equations of motion are
\begin{align}
	\partial_t^2 \phi_a &= \nabla^2 \phi_a
	- e^2 A_i A_i \phi_a - 2e \epsilon_{ab}\partial_i \phi_b A_i - e 
	\epsilon_{ab} \phi_b \Gamma \nonumber \\
	&	- \lambda (\phi_b\phi_b- \eta^2)\phi_a \label{phieq}  \\
	\partial_t F_{0i} &= \nabla^2 A_i - \partial_i \Gamma +  e 
	(\epsilon_{ab} \phi_a \partial_i \phi_b + e A_i \phi_a \phi_a) \label{Aeq}\\ 
	\partial_t \Gamma & = \partial_i F_{0i} - g_p^2 [ \partial_i 
	F_{0i}  + e \epsilon_{ab} \phi_a \partial_t \phi_b ],
\label{Gammaeq}
\end{align}
where $a=1,2$, $\epsilon_{ab}$ is the Levi-Civita tensor with 
$\epsilon_{12} = 1$, $F_{0i} = \partial_t A_i$ in the temporal gauge, 
$\lambda$ and $e$ are coupling constants, $\Gamma \equiv \partial_i A_i$,
and $g_p^2$ is a parameter introduced for numerical stability \cite{Vachaspati:2016abz}. 
The solution for a 
topologically stable 
straight string along the $z-$axis is \cite{Nielsen:1973cs}
\be
\phi = \eta f(r) e^{i\theta}, \ \ 
A_i = v(r) \epsilon_{ij} \frac{x^j}{r^2} \ \ (i,j=1,2),
\label{stringsoln}
\ee
where $r=\sqrt{x^2+y^2}$, $\theta=\tan^{-1}(y/x)$, and
$f(r)$ and $v(r)$ are 
profile functions that vanish
at the origin and asymptote to 1, respectively.
The string energy per unit length (also its tension) is given by
$\mu = \pi \eta^2 F(\beta )$
where $\beta \equiv 2\lambda/ e^2$ and $F$ is a numerically
determined function such that $F(1)=1$. 
We will only consider $\beta = 1$ corresponding to the Bogomol'nyi-Prasad-Sommerfield (BPS) limit 
\cite{Bogomolny:1975de,Prasad:1975kr}) where $\mu=\pi \eta^2$ and the scalar mass, $m_S = \sqrt{2\lambda} \eta$, equals the vector mass, $m_V =  e \eta$.

\begin{figure}[tbp]
      \includegraphics[width=0.23\textwidth,angle=0]{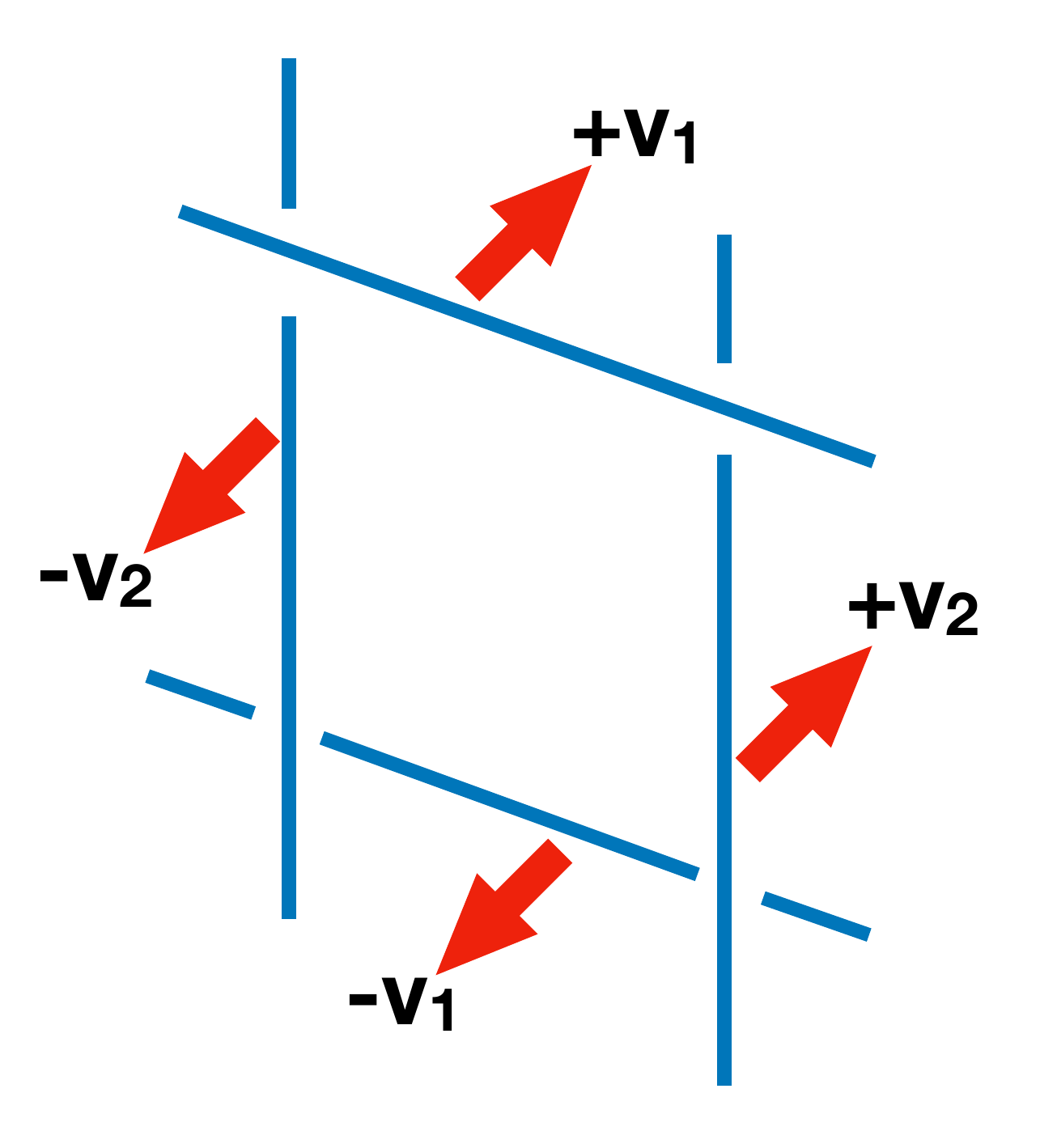}
  \caption{Schematic representation of the initial configuration. Four straight strings are
  set up with velocities as shown. The strings intersect and reconnect
  to produce a central loop and also a second ``outer'' loop because
  of periodic boundary conditions. These loops then oscillate and shrink
  without interacting with each other. By choosing the spacing of the
  initial strings, we can produce loops of different sizes.}
\label{schematics}
\end{figure}

Our aim is to produce a loop as might be produced in a cosmological
setting and then to evolve it. For this purpose, we set up initial conditions
with four straight strings that are moving with velocities $\pm \mathbf{v_1}$ and 
$\pm \mathbf{v_2}$ as shown schematically in Fig.~\ref{schematics}. 
The four strings then collide to form a loop with a stationary center of mass and a non-zero angular momentum.
The latter is essential to prevent the loop from simply collapsing to a double  line.
Preparing this initial configuration starts with taking the string solution of
Eq.~\eqref{stringsoln} oriented along a given direction, boosting it to a suitable velocity, and
gauge transforming the boosted solution back in to
the temporal gauge. Then the four string solutions have
to be patched together in a simulation box with periodic boundaries.
Further details are provided in the Supplemental Material section.

Cosmological strings are expected to be mildly relativistic and we choose $|\mathbf{v}_1|=0.6$ and 
$|\mathbf{v}_2|=0.33$. The directions are taken to be $({\hat v}_1)_x=0.4$, 
$({\hat v}_1)_y=\sqrt{1-0.4^2} \approx 0.92$ for the two strings oriented along the $z$-axis and 
$({\hat v}_2)_z=0.4$, $({\hat v}_2)_y \approx 0.92$ for those along the $x$-axis. The string 
velocities are approximately aligned along the $y$-axis, but not exactly, to avoid overly symmetrical 
loops that tend to pass through a double line configuration and collapse prematurely.
We have experimented with a wide range of initial velocities and our main conclusions are independent 
of the particular choices of these parameters.

Given the initial conditions for fields $\phi$, $A_\mu$,
we evolved them using the discretized version of Eqs.~\eqref{phieq}-\eqref{Gammaeq} with $e=1$, $\lambda = 1/2$, $\eta = 1$ and $g_p^2 =0.75$. We used the explicit Crank-Nicholson algorithm with two iterations for the evolution \cite{Teukolsky:1999rm} and 
periodic boundary conditions. We tried different lattice spacings to 
study the effects of numerical resolution. The initial string spacing was set to
a fixed fraction of the simulation box size so that smaller loops ran in a smaller box,
with less computational cost. 

Because of periodic boundary conditions, the reconnection of four strings produces two loops -- the central loop in the middle of the box shown in Fig.~\ref{schematics}, and an ``outer'' loop formed from the ``fragments'' in the corners of the box. The two loops then oscillate and decay without intersecting each other. We track the loop energy by summing the energy density
in the ``core'' of the string. The energy density is given by
\be
{\cal E} = \frac{1}{2} |D_0 \phi |^2+\frac{1}{2} |D_i \phi |^2 + \frac{1}{2} 
(\mathbf{E}^2 +\mathbf{B}^2) + \frac{\lambda}{4} (|\phi|^2-\eta^2)^2
\ee
where $\mathbf{E}$ and $\mathbf{B}$ are the electric and magnetic field vectors, with their components defined as $E_i = F_{0i}$ and $B_i = - 
\frac{1}{2} \epsilon_{ijk} F_{jk}$. We define the string core to be the cells where the magnitude of the 
scalar field, $|\phi |$, is less than $0.9\eta$. 

\begin{figure}[tbp]
     \includegraphics[width=0.5\textwidth,angle=0]{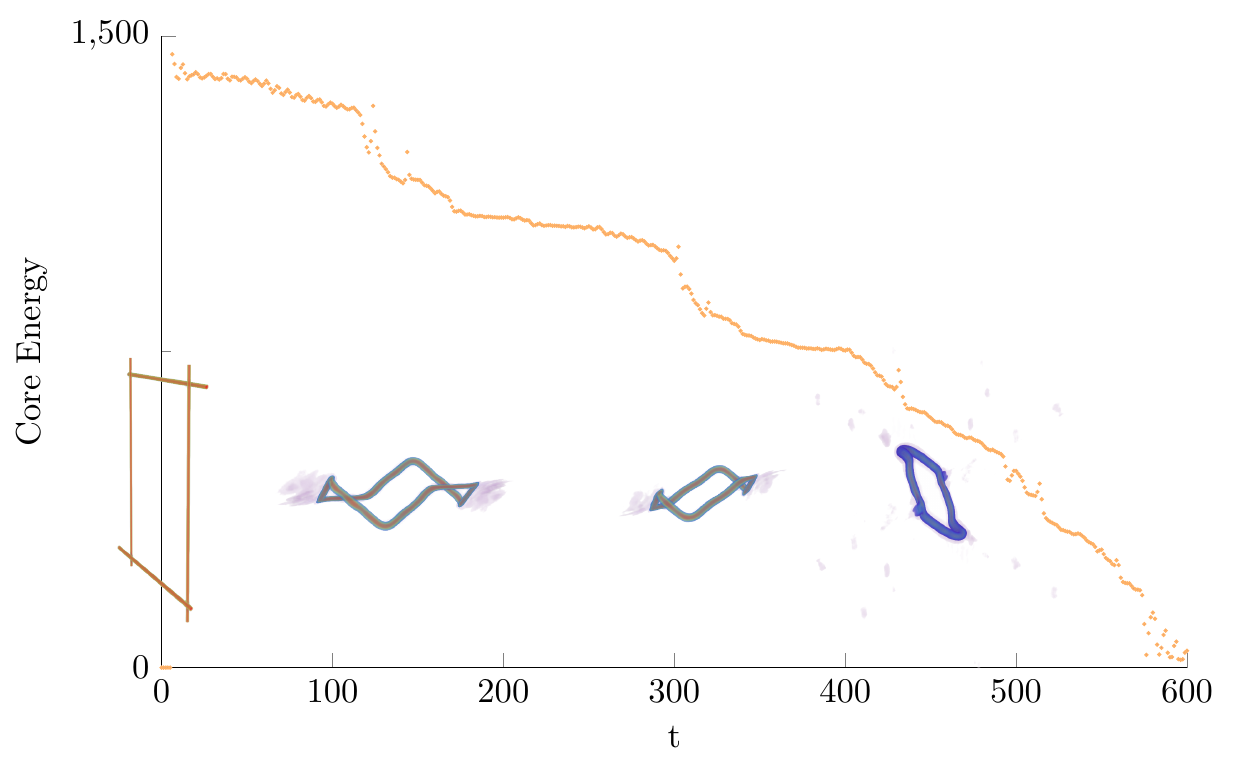}
  \caption{Energy of a loop with the initial size of $390$ lattice spacings plotted vs time. Overlaid on the plot are snapshots of the loop as it goes through phases of rapid radiation discharge due to smoothening of kinks. The animation showing the
 evolution of this loop can be found at \cite{movie}.}
\label{snapshots}
\end{figure}

In Fig.~\ref{snapshots} we plot the loop energy vs time for a simulation on a $600^3$ lattice with 
 $\Delta x=0.25$, where the initial size of the loop is $390$ lattice spacings. (The animation of the loop evolution 
 can be found at \cite{movie}.) The plot suggests episodic radiation,
 with the overlaid snapshots showing the representative ``events'' leading to drops in the loop energy. 
Straight strings do not radiate as they correspond to a boosted string solution. The kinks on the loop, formed at the intercommutation of the straight strings, also propagate with minimal energy loss. We find that noticeable radiation is produced when kinks collide.
Also, as the kinks smooth out, there are episodes of large radiation which may be due to the formation of 
weak cusps. Particle radiation from cusps was studied in Ref.~\cite{Olum:1998ag} where it was found
that the energy emission from a cusp leads to the formation of kinks and to weak cusps in
subsequent loop oscillations. This pattern of episodic radiation from kink collisions and weak cusps, with 
relatively minor energy loss in between these events, is common to all loop simulations we have performed.

\begin{table}[tbp]
\begin{tabular}{|c|c|c|c|c|}
\hline
Lattice size & Inner loop & Outer loop  \\
\hline
$400^3$ & 140 & 260 \\
$600^3$ & 210 & 390 \\
$800^3$ & 280 & 520 \\
$1200^3$ & 420 & 780\\
\hline
\end{tabular}
\caption{Loop sizes in lattice units for each of the runs.}
\label{runsTable}
\end{table}

\begin{figure}[tbp]
      \includegraphics[width=0.45\textwidth,angle=0]{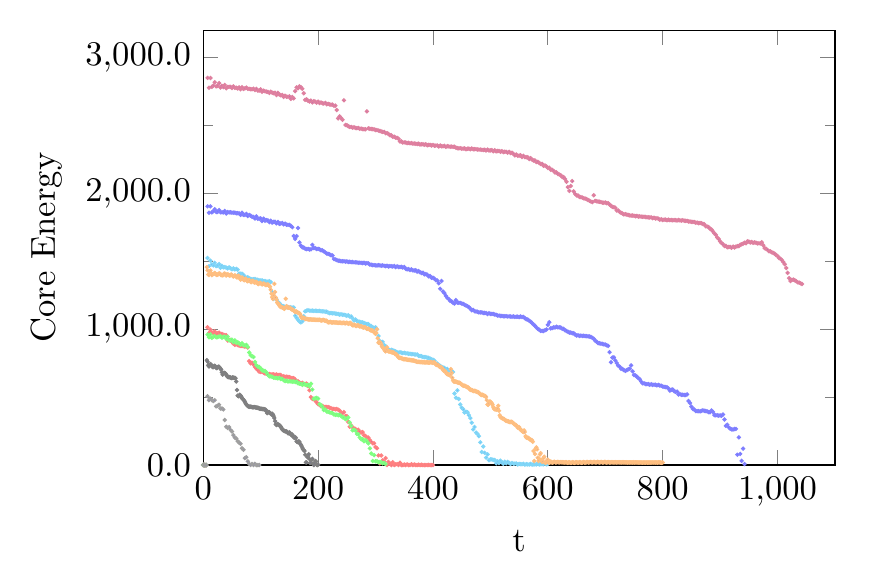}
  \caption{Loop energy vs. time for 8 different loops in 4 separate runs at $\Delta x=0.25$ resolution.}
\label{Evst}
\end{figure}

\begin{figure}[tbp]
      \includegraphics[width=0.4\textwidth,angle=0]{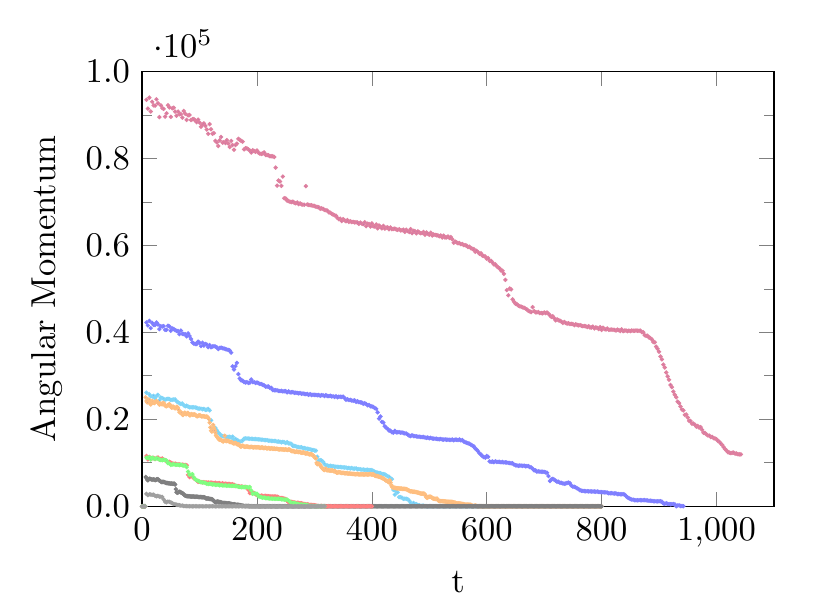}
  \caption{Loop angular momentum vs. time for 8 different loops in 4 separate runs at $\Delta x=0.25$ resolution.}
\label{AMvst}
\end{figure}

To obtain a quantitative measure of the scaling of the loop half-life with its size, we have runs simulations for 4 
different box sizes yielding 8 loops given in Table~\ref{runsTable}. (Two loops from different runs are almost the 
same length and provide 
a check on our simulation.) Fig.~\ref{Evst} shows the loop energy versus time for the 8 loops. 
As the loops evolve, they also shed their angular momentum, defined as
\begin{align}
	L_i \equiv \epsilon_{ijk} \int_{\rm string\, core} \hskip -1.0 cm d^3x \, x_j [ &  
	-\frac{1}{2}((D_0 \phi)(D_k \phi)^* + (D_0 \phi)^*(D_k \phi)) \nonumber 
	\\ & + \epsilon_{klm} E_l B_m].
\end{align}
In Fig.~\ref{AMvst} we plot $|{\mathbf L}|$ vs time and also see episodic decay.

\begin{figure*}[tbp]
     \includegraphics[width=0.4\textwidth,angle=0]{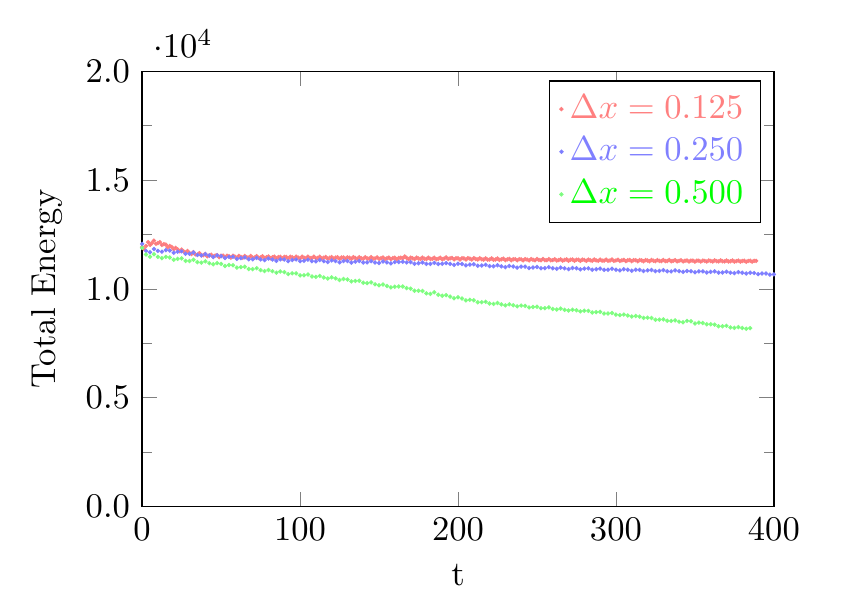}
      \includegraphics[width=0.43\textwidth,angle=0]{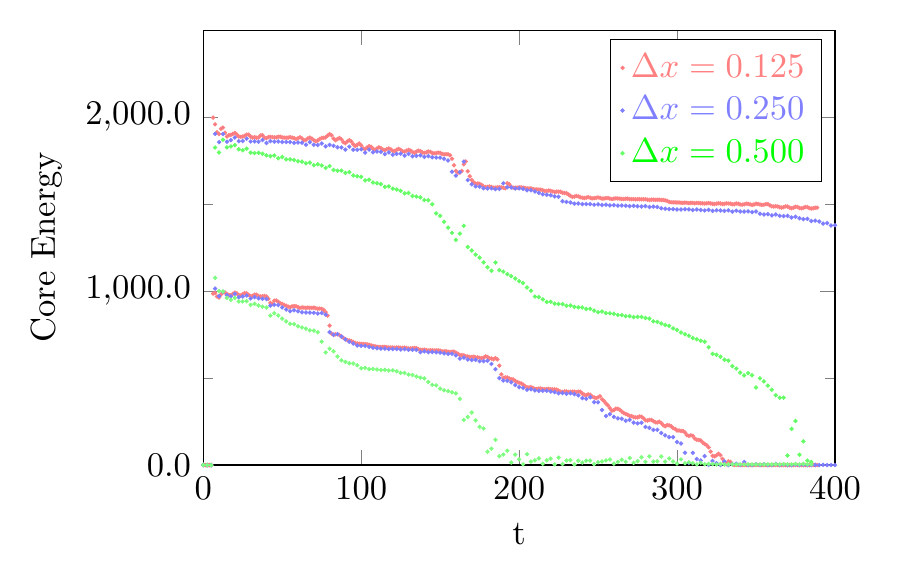}
  \caption{Comparison of runs with different lattice resolution $\Delta x=0.125,\, 0.25,\, 0.5$
  on lattices of size 1600, 800 and 400, respectively, corresponding to a fixed physical
  lattice length of 100. The left panel shows the total energy in our simulation box and the right panel shows the 
  evolution of the energy in the two loops in the box. The plots show convergence at higher 
  resolution and that $\Delta x=0.25$ offers a good compromise between accuracy and speed.}
\label{case4v06ResolutionTest}
\end{figure*}

We have run our simulations for a few different values of the lattice spacing, $\Delta x$, and found that the 
results are sensitive to the resolution. For example, as shown in Fig.~\ref{case4v06ResolutionTest}, the
total energy in the simulation box over the entire run is conserved only at $\sim 33\%$ level
when $\Delta x = 0.50$. For $\eta=1$, $e=1$, $\lambda=1/2$, the string
width is $\sim 1$. Therefore, with $\Delta x=0.5$ we only have a few lattice points within the width of
the string. Using $\Delta x=0.25$ improves the conservation to $\sim 5\%$ level
and agrees well with the much more computationally expensive run with $\Delta x = 0.125$.
The choice of $\Delta x$ makes an important difference in the lifetime of the loop, as is clear
from the right panel of Fig.~\ref{case4v06ResolutionTest}. Loops live longer in simulations 
with better resolution. From the animations, we see that the shorter loops live for
about one oscillation period while the larger loops survive for several oscillation periods.
(There is ambiguity in defining an oscillation
period since the length of the loop and hence its oscillation period is changing relatively
rapidly during the simulation.)

\begin{figure}[tbp]
      \includegraphics[width=0.4\textwidth,angle=0]{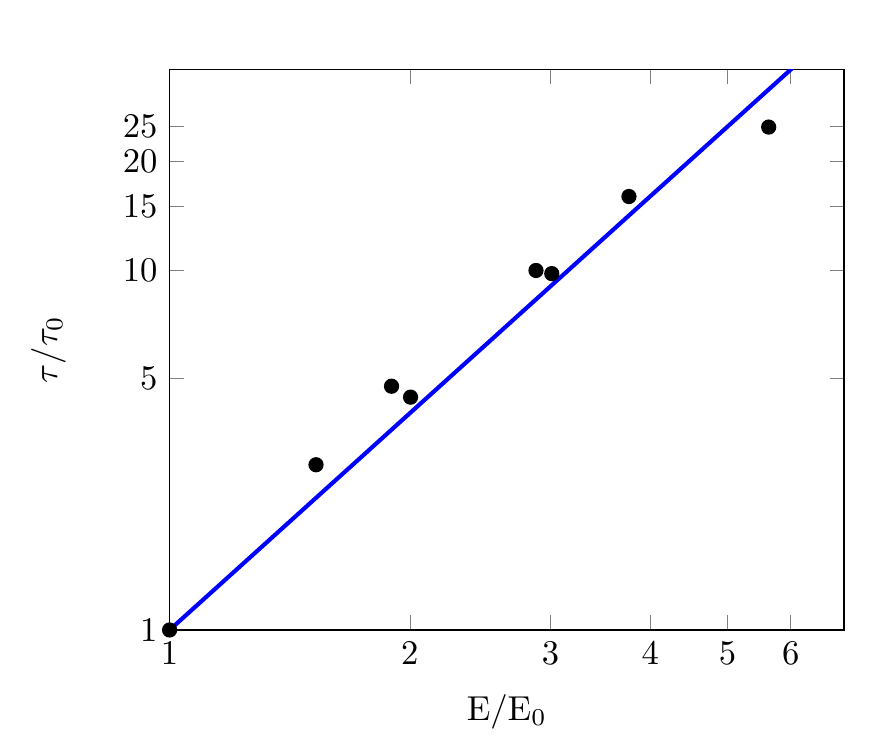}
  \caption{Plot of the loop half-life versus initial energy (proportional to the
initial length) in the loop. They are normalized by the initial half-life of the smallest loop, 
$\tau_0$, and and its energy, $E_0$. The straight line fit shows that
$\tau \propto L^2$ where $L$ is the initial length of the loop.}
\label{decaytime}
\end{figure}

The longest loop we are able to simulate has energy $\sim 3\times 10^3$, which corresponds
to length $L \sim 10^3 w$ where $w$ is the width of the string. In cosmology we are interested
in loops of length comparable to the cosmic horizon, which is orders of 
magnitude larger than the thickness of the string, perhaps by a factor $\sim 10^{60}$.
So we need to extrapolate our results to larger lengths. For this purpose we calculate the
half-life, $\tau$, {\it i.e.} the time it takes the loop to lose half its
initial energy. In Fig.~\ref{decaytime} we plot $\tau/\tau_0$, where $\tau_0 = 41.5/\eta$ is the half-life of the smallest loop
in our simulations, versus the initial energy normalized by that of the smallest loop (denoted $E_0 = 506 \eta$). We find a power law fit,
\be
\tau = \tau_0 \left ( \frac{E}{E_0} \right )^p = \frac{1.6 \times 10^{-3}}{\eta} (\eta L)^p, \ \ p \approx 2
\label{tauL}
\ee
where we have reinserted dimensional factors of $\eta$. 

The $L^2$ scaling in \eqref{tauL} can be understood as following from radiation 
being due to episodes involving a fixed number of features (kinks and weak cusps) on the loop, 
with the power emitted in a given episode (a kink collision or a weak cusp) being independent of $L$.
(Note that the size of the steps seen in Fig.~\ref{Evst} is similar for different loops).
If $\nu$ denotes the number of episodes per period and each episode radiates
energy $\epsilon$ on average, the energy lost per unit time is 
\be
{\dot E} \sim - \frac{\nu \epsilon}{L} \sim - \frac{ \mu \nu \epsilon}{E}.
\label{Edot}
\ee
Integration of this equation gives a lifetime 
\be
\tau \sim \frac{E^2}{\mu\nu \epsilon} \sim \frac{\mu L^2}{\nu\epsilon}
\label{tauEstimate}
\ee
in agreement with the $L^2$ scaling in \eqref{tauL}.

The particle radiation rate (\ref{Edot}) is to be contrasted with ${\dot E} \sim \nu G\mu^2$ expected
due to gravitational wave radiation from $\nu$ radiation episodes involving kinks and 
cusps \cite{Vachaspati:1984gt,Garfinkle:1987yw,Binetruy:2009vt}. Note that the rate of
energy loss to gravitational radiation is not suppressed by a factor of $L$ as is the case
for particle radiation in \eqref{Edot}. This is because, for example, a cusp on a loop that
is twice as large is also twice as large, and the gravitational energy emitted by a single cusp
is proportional to $L$. 
Then the lifetime of the loop due to gravitational radiation is
\be
\tau_g \sim \frac{L}{\nu G\mu}.
\label{eq:taug}
\ee
Comparing this to \eqref{tauEstimate} allows us to derive a criterion for when the gravitational radiation 
is more important than particle radiation, namely, when
\be
\tau_g < \tau \ \ \Rightarrow \ \ L \gtrsim \frac{\epsilon}{G\mu^2} \sim \frac{w}{G\mu}
\label{condition}
\ee 
where $w$ is the string thickness and we estimate $\epsilon \sim \sqrt{\mu}$, 
{\it i.e.} the particle energy emitted in an 
episode is comparable to the energy scale of the string,
and $l_P \sim 10^{-33} ~{\rm cm}$ is the Planck length.
Note that $\nu$ has canceled out in \eqref{condition}.
Therefore, even if there are more episodes on larger loops, gravitational
radiation still dominates over particle radiation if  \eqref{condition} is satisfied.

With $L \sim 10^{27}~{\rm cm}$ we find that gravitational radiation is less
important than particle radiation if $G\mu \lesssim 10^{-40}$, corresponding to 
$\eta \sim 100~{\rm MeV}$ or the QCD scale. Hence particle radiation could be
the main decay mechanism for strings formed below the QCD scale
but the dynamics of strings formed at such low energies is expected
to be dominated by friction with the ambient medium~\cite{Vilenkin:2000jqa}.

Alternately, for strings close to the current bound on the string tension,
$G\mu \approx 10^{-11}$, Eq.~\eqref{condition} implies that particle radiation
will only be important for loops that are very small, $L < 10^{-17}~{\rm cm}$.
Most of the radiation from such a network of strings will be in gravitational waves.

We would like to point out some caveats to the above discussion. The first caveat is that
the long strings in our initial conditions are straight and smooth. If these strings started
out with structure (perhaps as shallow kinks) on them, as has been suggested in
Ref.~\cite{Hindmarsh:2008dw},
the number of radiative episodes would be larger, and
both the particle and gravitational radiation would be larger. This would not change
the relative importance of particle and gravitational radiation but it would mean that
the loop decays faster. A second caveat is that our loops only contain kinks and no
cusps. It is known from Ref.~\cite{Olum:1998ag} 
that the particle radiation loss from a cusp is proportional
to $\sqrt{L}$ and this does not agree with our model where each episode emits radiation
that is independent of $L$. However, once the cusp radiates, it forms two kinks that then
propagate, radiate and smooth out to some extent. In the next oscillation, the cusp is
weaker and the energy radiated will not be proportional to $\sqrt{L}$, instead it will
be proportional to some power of $L$ smaller than 1/2. Thus with cusps we expect
that the effective value of $p$ will satisfy $1< p < 3/2$, and Eq.~\eqref{condition}
will get modified. Even then there will be a critical loop length such that gravitational
emission dominates over particle radiation for larger loops. 
A third caveat is that since our initial strings were straight, there 
was no radiation while the kinks propagate on the straight segments. If, however,
the segments are curved, there will be some radiation even from a propagating
kink. This radiation would not be episodic but it would be suppressed by the
curvature of the segment, expected to be suppressed by the loop size divided 
by the cosmic horizon scale.

To summarize, we have studied the formation and evolution of cosmic string loops 
in field theory and estimated their lifetimes. We find that the lifetime of the loops is 
very sensitive to the resolution used in their numerical evolution. With insufficient resolution, 
the loops collapse within one oscillation period. At higher resolution, the loops survive for a few oscillation
periods and we observe that their lifetime grows as $L^2$. We can explain this
growth in terms of episodic particle radiation. When compared to gravitational
energy losses, we find that gravitational radiation dominates for loops 
that are larger than a critical length (see Eq.~\eqref{condition}).

\acknowledgements

We thank the Lorentz Center for hosting the Topological Defects workshop where
we had the opportunity to discuss these results with the participants, 
especially Jose Blanco-Pillado, Mark Hindmarsh, Ken Olum, Paul Shellard and 
Daniele Steer. DM and LP are supported in part by the National Sciences 
and Engineering Research Council (NSERC) of Canada. AS and TV are supported by 
the U.S.  Department of Energy, Office of High Energy Physics, under Award No.  
DE-SC0019470 at Arizona State University.
This research was enabled in part by 
support provided by WestGrid \cite{westgrid} and Compute Canada \cite{compute}. 
The bulk of the computations were performed on the Agave and 
Stampede2 clusters at Arizona State University and The University of Texas at 
Austin, respectively.

\appendix
\section{Supplemental Material}

In this Section, we describe the steps involved in setting up the initial configuration of fields in our simulation.
First, we numerically find a solution for a static, infinite, straight string by
substituting the Nielsen-Olesen ansatz (\ref{stringsoln}) into the equations of motion and solving for the profile functions $f(r)$ and $v(r)$ using the relaxation method. We then use cubic spline interpolation to obtain smooth functions $f(r)$ and $v(r)$ and their derivatives.

To build a cosmic string loop we consider a string-antistring pair along the z-direction and
another along the x-direction in a simulation box with periodic boundary conditions (PBC).
We then Lorentz boost the strings and antistrings so that they are moving towards each other
as shown in Fig.~\ref{schematics}.
Since the evolution equations (\ref{Gammaeq}) are in the temporal gauge, we must gauge transform the boosted string solutions to set the temporal component of the gauge field to zero. Namely, we find a gauge transform $U=e^{ie\xi}$ such that 
\begin{eqnarray}
\label{eq:dtxi}
{A_0} &=& \bar{A}_{0} + \frac{i}{e} U \partial_{t} U^* = \bar{A}_{0} + \partial_{t} \xi = 0,\\
{A_i} &=& \bar{A}_{i} + \frac{i}{e} U \partial_{i} U^* = \bar{A}_{i} + \partial_{i} \xi,
\end{eqnarray}
where $A_\mu$ is in the temporal gauge and $\bar{A}_{\mu}$ is the field after the Lorentz boost. 

From (\ref{eq:dtxi}), we have $\partial_{t} \xi = - \bar{A}_{0}$ and $\xi$ can be evaluated as
\begin{equation}
\xi = \int_{0}^{t} d\tau \bar{A}_{0}.
\end{equation}
At initial time $t = 0$, this gives $\xi = 0$. Similarly $\left.\partial_i \xi\right|_{t=0}=0$.
Hence, at the initial time, we have
\begin{eqnarray}
{A_0} &=& 0,\\
{A_i} &=& \bar{A}_{i} + \left.\partial_i \xi\right|_{t=0} = \bar{A}_{i},
\end{eqnarray}
where all functions are evaluated at $t=0$. Note that the initial value of the scalar field is unaffected by the gauge transformation since $\exp(ie\xi)=1$ when $\xi=0$.

To solve the equations of motion we also need $\partial_{t} {A_\mu}$ and $\partial_t {\phi}$ at the initial time. We have
\begin{eqnarray}
\left. \partial_{t} {A_\mu} \right|_{t=0} &=& \partial_{t} \bar{A}_\mu  + \partial_{t}\partial_\mu \xi = \partial_{t} \bar{A}_\mu  - \partial_{\mu} \bar{A}_0 \\
\left. \partial_t {\phi}\right|_{t=0} &=& \partial_t \bar{\phi} - ie \bar{A}_0 \bar{\phi}.
\end{eqnarray}
where all functions are evaluated at $t=0$. 

To combine the string and anti-string solutions, we take the ansatz given by \cite{Vilenkin:2000jqa}
\begin{eqnarray}
\phi_{s\bar{s}} &=& \frac{ \phi_s\phi_{\bar{s}}}{\eta}= \frac{|\phi_s| |\phi_{\bar{s}}|}{\eta} e^{i (\theta_s - \theta_{\bar{s}})}, \\
A_{s\bar{s}} &=& A_{s} - A_{\bar{s}}.
\label{eq:ansatz}
\end{eqnarray}
To be consistent with the PBC, the phase of $\phi_{s\bar{s}}$ must approach zero at the boundaries of the box. While in (\ref{eq:ansatz}) the phase approaches zero asymptotically at infinity, it does not do so in a finite simulation box. Thus, we modified the ansatz to make the phase approach zero faster:
\begin{eqnarray}
\nonumber
 \phi_{s\bar{s},\text{mod}}
=  \frac{|\phi_s| |\phi_{\bar{s}}|}{\eta}  e^{i(\theta_s - \theta_{\bar{s}})[1 - \tanh{\left(\omega\left(\rho - \text{L}/2\right)\right)}]/2},
\end{eqnarray}
where $\omega$, taken to be $0.5$, is a parameter that determines how quickly the phase approaches to 0 at the boundaries, and L is the size of the box. Finally, the scalar and gauge fields of the two sets of a parallel string-antistring pair are given as
\begin{eqnarray}
\phi &=& \frac{ \phi_{s\bar{s},1}\phi_{s\bar{s},2}}{\eta}, \\
A &=& A_{s\bar{s},1} + A_{s\bar{s},2},
\end{eqnarray}
where $\phi_{s\bar{s},1},A_{s\bar{s},1}$ and $\phi_{s\bar{s},2},A_{s\bar{s},2}$ are the scalar and gauge fields of the first and second string - anti-string pairs, respectively.

\bibstyle{aps}


\begin{thebibliography}{29}
\expandafter\ifx\csname natexlab\endcsname\relax\def\natexlab#1{#1}\fi
\expandafter\ifx\csname bibnamefont\endcsname\relax
  \def\bibnamefont#1{#1}\fi
\expandafter\ifx\csname bibfnamefont\endcsname\relax
  \def\bibfnamefont#1{#1}\fi
\expandafter\ifx\csname citenamefont\endcsname\relax
  \def\citenamefont#1{#1}\fi
\expandafter\ifx\csname url\endcsname\relax
  \def\url#1{\texttt{#1}}\fi
\expandafter\ifx\csname urlprefix\endcsname\relax\def\urlprefix{URL }\fi
\providecommand{\bibinfo}[2]{#2}
\providecommand{\eprint}[2][]{\url{#2}}

\bibitem[{\citenamefont{Vilenkin and Shellard}(2000)}]{Vilenkin:2000jqa}
\bibinfo{author}{\bibfnamefont{A.}~\bibnamefont{Vilenkin}} \bibnamefont{and}
  \bibinfo{author}{\bibfnamefont{E.~P.~S.} \bibnamefont{Shellard}},
  \emph{\bibinfo{title}{{Cosmic Strings and Other Topological Defects}}}
  (\bibinfo{publisher}{Cambridge University Press}, \bibinfo{year}{2000}), ISBN
  \bibinfo{isbn}{9780521654760},
  \urlprefix\url{http://www.cambridge.org/mw/academic/subjects/physics/theoretical-physics-and-mathematical-physics/cosmic-strings-and-other-topological-defects?format=PB}.

\bibitem[{\citenamefont{Copeland et~al.}(2011)\citenamefont{Copeland, Pogosian,
  and Vachaspati}}]{Copeland:2011dx}
\bibinfo{author}{\bibfnamefont{E.~J.} \bibnamefont{Copeland}},
  \bibinfo{author}{\bibfnamefont{L.}~\bibnamefont{Pogosian}}, \bibnamefont{and}
  \bibinfo{author}{\bibfnamefont{T.}~\bibnamefont{Vachaspati}},
  \bibinfo{journal}{Class. Quant. Grav.} \textbf{\bibinfo{volume}{28}},
  \bibinfo{pages}{204009} (\bibinfo{year}{2011}), \eprint{1105.0207}.

\bibitem[{\citenamefont{Abbott et~al.}(2018)}]{Abbott:2017mem}
\bibinfo{author}{\bibfnamefont{B.}~\bibnamefont{Abbott}} \bibnamefont{et~al.}
  (\bibinfo{collaboration}{LIGO Scientific, Virgo}), \bibinfo{journal}{Phys.
  Rev.} \textbf{\bibinfo{volume}{D97}}, \bibinfo{pages}{102002}
  (\bibinfo{year}{2018}), \eprint{1712.01168}.

\bibitem[{\citenamefont{Lasky et~al.}(2016)}]{Lasky:2015lej}
\bibinfo{author}{\bibfnamefont{P.~D.} \bibnamefont{Lasky}}
  \bibnamefont{et~al.}, \bibinfo{journal}{Phys. Rev.}
  \textbf{\bibinfo{volume}{X6}}, \bibinfo{pages}{011035}
  (\bibinfo{year}{2016}), \eprint{1511.05994}.

\bibitem[{\citenamefont{Blanco-Pillado
  et~al.}(2018)\citenamefont{Blanco-Pillado, Olum, and
  Siemens}}]{Blanco-Pillado:2017rnf}
\bibinfo{author}{\bibfnamefont{J.~J.} \bibnamefont{Blanco-Pillado}},
  \bibinfo{author}{\bibfnamefont{K.~D.} \bibnamefont{Olum}}, \bibnamefont{and}
  \bibinfo{author}{\bibfnamefont{X.}~\bibnamefont{Siemens}},
  \bibinfo{journal}{Phys. Lett.} \textbf{\bibinfo{volume}{B778}},
  \bibinfo{pages}{392} (\bibinfo{year}{2018}), \eprint{1709.02434}.

\bibitem[{\citenamefont{Albrecht and Turok}(1985)}]{Albrecht:1984xv}
\bibinfo{author}{\bibfnamefont{A.}~\bibnamefont{Albrecht}} \bibnamefont{and}
  \bibinfo{author}{\bibfnamefont{N.}~\bibnamefont{Turok}},
  \bibinfo{journal}{Phys. Rev. Lett.} \textbf{\bibinfo{volume}{54}},
  \bibinfo{pages}{1868} (\bibinfo{year}{1985}).

\bibitem[{\citenamefont{Bennett and Bouchet}(1990)}]{Bennett:1989yp}
\bibinfo{author}{\bibfnamefont{D.~P.} \bibnamefont{Bennett}} \bibnamefont{and}
  \bibinfo{author}{\bibfnamefont{F.~R.} \bibnamefont{Bouchet}},
  \bibinfo{journal}{Phys. Rev.} \textbf{\bibinfo{volume}{D41}},
  \bibinfo{pages}{2408} (\bibinfo{year}{1990}).

\bibitem[{\citenamefont{Allen and Shellard}(1990)}]{Allen:1990tv}
\bibinfo{author}{\bibfnamefont{B.}~\bibnamefont{Allen}} \bibnamefont{and}
  \bibinfo{author}{\bibfnamefont{E.~P.~S.} \bibnamefont{Shellard}},
  \bibinfo{journal}{Phys. Rev. Lett.} \textbf{\bibinfo{volume}{64}},
  \bibinfo{pages}{119} (\bibinfo{year}{1990}).

\bibitem[{\citenamefont{Blanco-Pillado
  et~al.}(2011)\citenamefont{Blanco-Pillado, Olum, and
  Shlaer}}]{BlancoPillado:2011dq}
\bibinfo{author}{\bibfnamefont{J.~J.} \bibnamefont{Blanco-Pillado}},
  \bibinfo{author}{\bibfnamefont{K.~D.} \bibnamefont{Olum}}, \bibnamefont{and}
  \bibinfo{author}{\bibfnamefont{B.}~\bibnamefont{Shlaer}},
  \bibinfo{journal}{Phys. Rev.} \textbf{\bibinfo{volume}{D83}},
  \bibinfo{pages}{083514} (\bibinfo{year}{2011}), \eprint{1101.5173}.

\bibitem[{\citenamefont{Lorenz et~al.}(2010)\citenamefont{Lorenz, Ringeval, and
  Sakellariadou}}]{Lorenz:2010sm}
\bibinfo{author}{\bibfnamefont{L.}~\bibnamefont{Lorenz}},
  \bibinfo{author}{\bibfnamefont{C.}~\bibnamefont{Ringeval}}, \bibnamefont{and}
  \bibinfo{author}{\bibfnamefont{M.}~\bibnamefont{Sakellariadou}},
  \bibinfo{journal}{JCAP} \textbf{\bibinfo{volume}{1010}}, \bibinfo{pages}{003}
  (\bibinfo{year}{2010}), \eprint{1006.0931}.

\bibitem[{\citenamefont{Hindmarsh et~al.}(2017)\citenamefont{Hindmarsh,
  Lizarraga, Urrestilla, Daverio, and Kunz}}]{Hindmarsh:2017qff}
\bibinfo{author}{\bibfnamefont{M.}~\bibnamefont{Hindmarsh}},
  \bibinfo{author}{\bibfnamefont{J.}~\bibnamefont{Lizarraga}},
  \bibinfo{author}{\bibfnamefont{J.}~\bibnamefont{Urrestilla}},
  \bibinfo{author}{\bibfnamefont{D.}~\bibnamefont{Daverio}}, \bibnamefont{and}
  \bibinfo{author}{\bibfnamefont{M.}~\bibnamefont{Kunz}},
  \bibinfo{journal}{Phys. Rev.} \textbf{\bibinfo{volume}{D96}},
  \bibinfo{pages}{023525} (\bibinfo{year}{2017}), \eprint{1703.06696}.

\bibitem[{\citenamefont{Vachaspati et~al.}(1984)\citenamefont{Vachaspati,
  Everett, and Vilenkin}}]{Vachaspati:1984yi}
\bibinfo{author}{\bibfnamefont{T.}~\bibnamefont{Vachaspati}},
  \bibinfo{author}{\bibfnamefont{A.~E.} \bibnamefont{Everett}},
  \bibnamefont{and} \bibinfo{author}{\bibfnamefont{A.}~\bibnamefont{Vilenkin}},
  \bibinfo{journal}{Phys. Rev.} \textbf{\bibinfo{volume}{D30}},
  \bibinfo{pages}{2046} (\bibinfo{year}{1984}).

\bibitem[{\citenamefont{Srednicki and Theisen}(1987)}]{Srednicki:1986xg}
\bibinfo{author}{\bibfnamefont{M.}~\bibnamefont{Srednicki}} \bibnamefont{and}
  \bibinfo{author}{\bibfnamefont{S.}~\bibnamefont{Theisen}},
  \bibinfo{journal}{Phys. Lett.} \textbf{\bibinfo{volume}{B189}},
  \bibinfo{pages}{397} (\bibinfo{year}{1987}).

\bibitem[{\citenamefont{Brandenberger}(1987)}]{Brandenberger:1986vj}
\bibinfo{author}{\bibfnamefont{R.~H.} \bibnamefont{Brandenberger}},
  \bibinfo{journal}{Nucl. Phys.} \textbf{\bibinfo{volume}{B293}},
  \bibinfo{pages}{812} (\bibinfo{year}{1987}).

\bibitem[{\citenamefont{Olum and Blanco-Pillado}(1999)}]{Olum:1998ag}
\bibinfo{author}{\bibfnamefont{K.~D.} \bibnamefont{Olum}} \bibnamefont{and}
  \bibinfo{author}{\bibfnamefont{J.~J.} \bibnamefont{Blanco-Pillado}},
  \bibinfo{journal}{Phys. Rev.} \textbf{\bibinfo{volume}{D60}},
  \bibinfo{pages}{023503} (\bibinfo{year}{1999}), \eprint{gr-qc/9812040}.

\bibitem[{\citenamefont{Olum and Blanco-Pillado}(2000)}]{Olum:1999sg}
\bibinfo{author}{\bibfnamefont{K.~D.} \bibnamefont{Olum}} \bibnamefont{and}
  \bibinfo{author}{\bibfnamefont{J.~J.} \bibnamefont{Blanco-Pillado}},
  \bibinfo{journal}{Phys. Rev. Lett.} \textbf{\bibinfo{volume}{84}},
  \bibinfo{pages}{4288} (\bibinfo{year}{2000}), \eprint{astro-ph/9910354}.

\bibitem[{\citenamefont{Martins et~al.}(2004)\citenamefont{Martins, Moore, and
  Shellard}}]{Martins:2003vd}
\bibinfo{author}{\bibfnamefont{C.~J. A.~P.} \bibnamefont{Martins}},
  \bibinfo{author}{\bibfnamefont{J.~N.} \bibnamefont{Moore}}, \bibnamefont{and}
  \bibinfo{author}{\bibfnamefont{E.~P.~S.} \bibnamefont{Shellard}},
  \bibinfo{journal}{Phys. Rev. Lett.} \textbf{\bibinfo{volume}{92}},
  \bibinfo{pages}{251601} (\bibinfo{year}{2004}), \eprint{hep-ph/0310255}.

\bibitem[{\citenamefont{Hindmarsh et~al.}(2009)\citenamefont{Hindmarsh,
  Stuckey, and Bevis}}]{Hindmarsh:2008dw}
\bibinfo{author}{\bibfnamefont{M.}~\bibnamefont{Hindmarsh}},
  \bibinfo{author}{\bibfnamefont{S.}~\bibnamefont{Stuckey}}, \bibnamefont{and}
  \bibinfo{author}{\bibfnamefont{N.}~\bibnamefont{Bevis}},
  \bibinfo{journal}{Phys. Rev.} \textbf{\bibinfo{volume}{D79}},
  \bibinfo{pages}{123504} (\bibinfo{year}{2009}), \eprint{0812.1929}.

\bibitem[{\citenamefont{Vachaspati}(2016)}]{Vachaspati:2016abz}
\bibinfo{author}{\bibfnamefont{T.}~\bibnamefont{Vachaspati}},
  \bibinfo{journal}{Phys. Rev. Lett.} \textbf{\bibinfo{volume}{117}},
  \bibinfo{pages}{181601} (\bibinfo{year}{2016}), \eprint{1607.07460}.

\bibitem[{\citenamefont{Nielsen and Olesen}(1973)}]{Nielsen:1973cs}
\bibinfo{author}{\bibfnamefont{H.~B.} \bibnamefont{Nielsen}} \bibnamefont{and}
  \bibinfo{author}{\bibfnamefont{P.}~\bibnamefont{Olesen}},
  \bibinfo{journal}{Nucl. Phys.} \textbf{\bibinfo{volume}{B61}},
  \bibinfo{pages}{45} (\bibinfo{year}{1973}), \bibinfo{note}{[,302(1973)]}.

\bibitem[{\citenamefont{Bogomolny}(1976)}]{Bogomolny:1975de}
\bibinfo{author}{\bibfnamefont{E.~B.} \bibnamefont{Bogomolny}},
  \bibinfo{journal}{Sov. J. Nucl. Phys.} \textbf{\bibinfo{volume}{24}},
  \bibinfo{pages}{449} (\bibinfo{year}{1976}), \bibinfo{note}{[Yad.
  Fiz.24,861(1976)]}.

\bibitem[{\citenamefont{Prasad and Sommerfield}(1975)}]{Prasad:1975kr}
\bibinfo{author}{\bibfnamefont{M.~K.} \bibnamefont{Prasad}} \bibnamefont{and}
  \bibinfo{author}{\bibfnamefont{C.~M.} \bibnamefont{Sommerfield}},
  \bibinfo{journal}{Phys. Rev. Lett.} \textbf{\bibinfo{volume}{35}},
  \bibinfo{pages}{760} (\bibinfo{year}{1975}).

\bibitem[{\citenamefont{Teukolsky}(2000)}]{Teukolsky:1999rm}
\bibinfo{author}{\bibfnamefont{S.~A.} \bibnamefont{Teukolsky}},
  \bibinfo{journal}{Phys. Rev.} \textbf{\bibinfo{volume}{D61}},
  \bibinfo{pages}{087501} (\bibinfo{year}{2000}), \eprint{gr-qc/9909026}.

\bibitem[{mov()}]{movie}
\urlprefix\url{https://ayushsaurabh.home.blog}.

\bibitem[{\citenamefont{Vachaspati and Vilenkin}(1985)}]{Vachaspati:1984gt}
\bibinfo{author}{\bibfnamefont{T.}~\bibnamefont{Vachaspati}} \bibnamefont{and}
  \bibinfo{author}{\bibfnamefont{A.}~\bibnamefont{Vilenkin}},
  \bibinfo{journal}{Phys. Rev.} \textbf{\bibinfo{volume}{D31}},
  \bibinfo{pages}{3052} (\bibinfo{year}{1985}).

\bibitem[{\citenamefont{Garfinkle and Vachaspati}(1987)}]{Garfinkle:1987yw}
\bibinfo{author}{\bibfnamefont{D.}~\bibnamefont{Garfinkle}} \bibnamefont{and}
  \bibinfo{author}{\bibfnamefont{T.}~\bibnamefont{Vachaspati}},
  \bibinfo{journal}{Phys. Rev.} \textbf{\bibinfo{volume}{D36}},
  \bibinfo{pages}{2229} (\bibinfo{year}{1987}).

\bibitem[{\citenamefont{Binetruy et~al.}(2009)\citenamefont{Binetruy, Bohe,
  Hertog, and Steer}}]{Binetruy:2009vt}
\bibinfo{author}{\bibfnamefont{P.}~\bibnamefont{Binetruy}},
  \bibinfo{author}{\bibfnamefont{A.}~\bibnamefont{Bohe}},
  \bibinfo{author}{\bibfnamefont{T.}~\bibnamefont{Hertog}}, \bibnamefont{and}
  \bibinfo{author}{\bibfnamefont{D.~A.} \bibnamefont{Steer}},
  \bibinfo{journal}{Phys. Rev.} \textbf{\bibinfo{volume}{D80}},
  \bibinfo{pages}{123510} (\bibinfo{year}{2009}), \eprint{0907.4522}.

\bibitem[{wes()}]{westgrid}
\urlprefix\url{http://www.westgrid.ca}.

\bibitem[{com()}]{compute}
\urlprefix\url{http://www.computecanada.ca}.

\end{thebibliography}
\end{document}